\documentclass[conference]{IEEEtran}
\ifCLASSINFOpdf
\else
\fi
\usepackage{amssymb}
\usepackage{graphicx}
\usepackage{graphics}
\usepackage[cmex10]{amsmath}
\usepackage{amssymb, amsfonts}
\usepackage{subfigure}
\usepackage{slashbox}
\usepackage{caption}
\usepackage[T1]{fontenc}
\usepackage[utf8]{inputenc}
\usepackage{graphics}                 
\usepackage{graphicx}
\usepackage[cmex10]{amsmath}
\usepackage{amssymb, amsfonts}
\usepackage{algorithmic}
\usepackage{algorithm}
\usepackage{verbatim}
\usepackage{url}
\usepackage{subfigure}
\usepackage{slashbox}
\usepackage{setspace}
\usepackage{multirow}
\usepackage{array}
\usepackage{hhline}

\graphicspath{{figures/}}

\usepackage{url}

\newcolumntype{P}[1]{>{\centering\arraybackslash}p{#1\textwidth}}

\begin{document}
%
\title{Tracking Topology Dynamicity for Link Prediction in Intermittently Connected Wireless Networks}

\author{\IEEEauthorblockN{Mohamed-Haykel Zayani, Vincent Gauthier, Ines Slama and Djamal Zeghlache}
\IEEEauthorblockA{Lab. CNRS SAMOVAR UMR 5157\\ Institut Mines-Telecom, Telecom SudParis\\
Evry, France\\\{mohamed-haykel.zayani, vincent.gauthier, ines.slama,
djamal.zeghlache\}@telecom-sudparis.eu} }

%


\IEEEoverridecommandlockouts
\IEEEpubid{\makebox[\columnwidth]{978-1-4577-1379-8/12/\$26.00~\copyright~2012
IEEE \hfill} \hspace{\columnsep}\makebox[\columnwidth]{ }}

\maketitle

\begin{abstract}
Through several studies, it has been highlighted that mobility
patterns in mobile networks are driven by human behaviors. This
effect has been particularly observed in intermittently connected
networks like DTN (Delay Tolerant Networks). Given that common
social intentions generate similar human behavior, it is relevant to
exploit this knowledge in the network protocols design, e.g. to
identify the closeness degree between two nodes. In this paper, we
propose a temporal link prediction technique for DTN which
quantifies the behavior similarity between each pair of nodes and
makes use of it to predict future links. We attest that the
tensor-based technique is effective for temporal link prediction
applied to the intermittently connected networks. The validity of
this method is proved when the prediction is made in a distributed
way (i.e. with local information) and its performance is compared to
well-known link prediction metrics proposed in the literature.
\end{abstract}




%
\IEEEpeerreviewmaketitle

\section{Introduction}

In recent years extensive research has addressed challenges and
problems raised in mobile, sparse and intermittently connected
networks (i.e. DTN). In this case, forwarding packets greatly
depends on the occurrence of contacts. Since the existence of links
is crucial to deliver data from a source to a destination, the
contacts and their properties emerge as a key issue in designing
efficient communication protocols \cite{Hossmann2010a}. Obviously,
the occurrence of links is determined by the behavior of the nodes
in the network \cite{Chaintreau07}. It has been widely shown in
\cite{Hsu2009a, Thakur2010} that human mobility is directed by
social intentions and reflects spatio-temporal regularity. A node
can follow other nodes to a specific location (spatial level) and
may bring out a behavior which may be regulated by a schedule
(temporal level). The social intentions that govern the behavior of
mobile users have also been observed through statistical analyses in
\cite{Chaintreau07,Karagiannis2007} by showing that the distribution
of inter-contact times follow a truncated power law.

With the intention of improving the performance of intermittently
connected wireless network protocols, it is paramount to track and
understand the behavior of the nodes. We aim to propose an approach
that analyzes the network statistics, quantifies the social
relationship between each pair of nodes and exploits this measure as
a score which indicates if a link would occur in the immediate
future. 

In this paper, we adapt a tensor-based link prediction algorithm
successfully designed for data-mining \cite{Acar2009,Dunlavy2011}.
Our proposal records the network structure for $T$ time periods and
predicts links occurrences for the $(T+1)^{th}$ period. This link
prediction technique is designed through two steps. First, tracking
time-dependent network snapshots in adjacency matrices which form a
tensor. Second, applying of the Katz measure \cite{Katz1953}
inspired from sociometry. To the best of our knowledge, this work is
the first to perform the prediction technique in a distributed way.
The assessment of its efficiency can be beneficial for the
improvement or the design of communication protocols in mobile,
sparse and intermittently connected networks.

The paper is organized as follows: Section 2 presents the related
work that highlights the growing interest to the social analysis and
justifies the recourse to the tensors and to the Katz measure to
perform predictions. In Section 3, we describe the two main steps
that characterize our proposal. Section 4 details simulation
scenarios used to evaluate the tensor-based prediction approach,
analyzes the obtained results and assesses its efficiency. Finally,
we conclude the paper in Section 5.

\section{Related Work}
Social Network Analysis (SNA) \cite{Wasserman1994, Katsaros2010a}
and ad-hoc networking have provided new perspectives for the design
of network protocols \cite{Hui2008, Daly2007, Hossmann2010}. These
protocols aim to exploit the social aspects and relationship
features between the nodes. Studies conducted in the field of SNA
have mainly focused on two kinds of concepts: the most well-known
centrality metrics suggested in
\cite{Wasserman1994,Page1999,Hwang2008,Chung1997} and the community
detection mechanisms proposed in
\cite{Bollobas1998,Newman2006,Palla2005,Wasserman1994}. From this
perspective, several works have tried to develop synthetic models
that aim to reproduce realistic moving patterns
\cite{Hsu2009a,Lee2009}. Nonetheless, the study done in
\cite{Hossmann2010a} has underlined the fact that synthetic models
cannot faithfully reproduce human behavior because these synthetic
models are only location-driven and they do not track social
intentions explicitly.

In their survey, Katsaros et al. \cite{Katsaros2010a} have
underlined the limits of these protocols when the network topology
is time-varying. The main drawback comes down to their inability to
model topology changes as they are based on graph theory tools. To
overcome this limit, tensor-based approaches have been used in some
works to build statistics on the behavior of nodes in wireless
networks over time as in \cite{Acer2010}. Thakur et al.
\cite{Thakur2010} have also developed a model using a collapsed
tensor that tracks user's location preferences (characterized by
probabilities) with a considered time granularity (week days for
example) in order to follow the emergence of ``behavior-aware" delay
tolerant networks closely.

As previously mentioned, tracking the social ties between network
entities enables us to understand how the network is structured.
Such tracking has led to the design of techniques for link
prediction. Link prediction in social networks has been addressed in
data mining applications as in \cite{Acar2009,Dunlavy2011}.
Concerning link prediction in community-based communication
networks, \cite{Wang2011} has highlighted salient measures that
allow link occurrence between network users to be predicted. These
metrics determine if a link occurrence is likely by quantifying the
degree of proximity of two nodes (Katz measure \cite{Katz1953}, the
number of common neighbors, Adamic-Adar measure \cite{Adamic2003},
Jaccard's coefficient \cite{Jaccard1901,Salton1986}, \ldots) or by
computing the similarity of their mobility patterns (spatial cosine
similarity, co-location rate, \ldots).

In this paper, we propose a link prediction technique that tracks
the temporal network topology evolution in a tensor and computes a
metric in order to characterize the social-based behavior similarity
of each pair of nodes. Some approaches have addressed the same
problem in data-mining in order to perform link prediction. Acar et
al. \cite{Acar2009} and Dunlavy et al. \cite{Dunlavy2011} have
provided detailed methods based on matrix and tensor factorizations
for link prediction in social networks such as the DBLP data set
\cite{DBLP}. These methods have been successfully applied to predict
a collaboration between two authors by recording the structure of
relationships over a tracking period. Moreover, they have
highlighted the use of the Katz measure \cite{Katz1953}, which can
be seen as a behavior similarity metric, by assigning a link
prediction score for each pair of nodes. The efficiency of the Katz
measure in link prediction has been also demonstrated in
\cite{Acar2009,Dunlavy2011,Wang2011,Liben-Nowell2007}.

\section{Description of the Tensor Based Prediction Method}
It has been highlighted that a human mobility pattern shows a high
degree of temporal and spatial regularity, and each individual is
characterized by a time-dependent mobility pattern and a trend to
return to preferred locations \cite{Chaintreau07, Hsu2009a,
Thakur2010}. In this paper, we propose an approach that aims to
exploit similar behavior of nodes in order to predict link
occurrence referring to the social closeness.

To quantify the social closeness between each pair of nodes in the
network, we use the Katz measure \cite{Katz1953} inspired by
sociometry. This measure aims at quantifying the social distance
between people inside a social network. We also need to use a
structure that records link occurrence between each pair of nodes
over a certain period of time in order to perform the similarity
measure computation. The records represent the network behavior
statistics in time and space. To this end, a third-order tensor is
considered. A tensor $\boldsymbol{\mathcal{Z}}$ consists of a set of
slices and each slice corresponds to an adjacency matrix of the
network tracked over a given period of time $p$. After the tracking
phase, we reduce the tensor into a matrix (or collapsed tensor)
which expresses the weight of each link according to its lifetime
and its recentness. A high weight value in this matrix denotes a
link whose corresponding nodes share a high degree of closeness. We
apply the Katz measure to the collapsed tensor to compute a matrix
of scores $\mathbf{S}$ that not only considers direct links but also
indirect links (multi-hop connections). The matrix of scores
expresses the degree of similarity of each pair of nodes according
to the spatial and the temporal levels. The higher the score is, the
better the similarity pattern gets. Therefore, two nodes that have a
high similarity score are more likely to have a common link in the
future.

\subsection{Notation}
Scalars are denoted by lowercase letters, e.g., $a$. Vectors are
denoted by boldface lowercase letters, e.g., $\bf{a}$. Matrices are
denoted by boldface capital letters, e.g., $\mathbf{A}$. The
$r^{th}$ column of a matrix $\mathbf{A}$ is denoted by $\bf{a_r}$.
Higher-order tensors are denoted by bold Euler script letters, e.g.,
$\boldsymbol{\mathcal{T}}$. The $n^{th}$ frontal slice of a tensor
$\boldsymbol{\mathcal{T}}$ is denoted $\mathbf{T_n}$. The $i^{th}$
entry of a vector $\bf{a}$ is denoted by $\bf{a}(i)$, element
$(i,j)$ of a matrix $\mathbf{A}$ is denoted by $\mathbf{A}(i,j)$,
and element $(i, j, k)$ of a third-order tensor
$\boldsymbol{\mathcal{T}}$ is denoted by $\mathbf{T_{i}}(j, k)$.

\subsection{Matrix of Scores Computation}
\begin{figure}[!tb]
    \centering
    \includegraphics[width=0.4\textwidth]{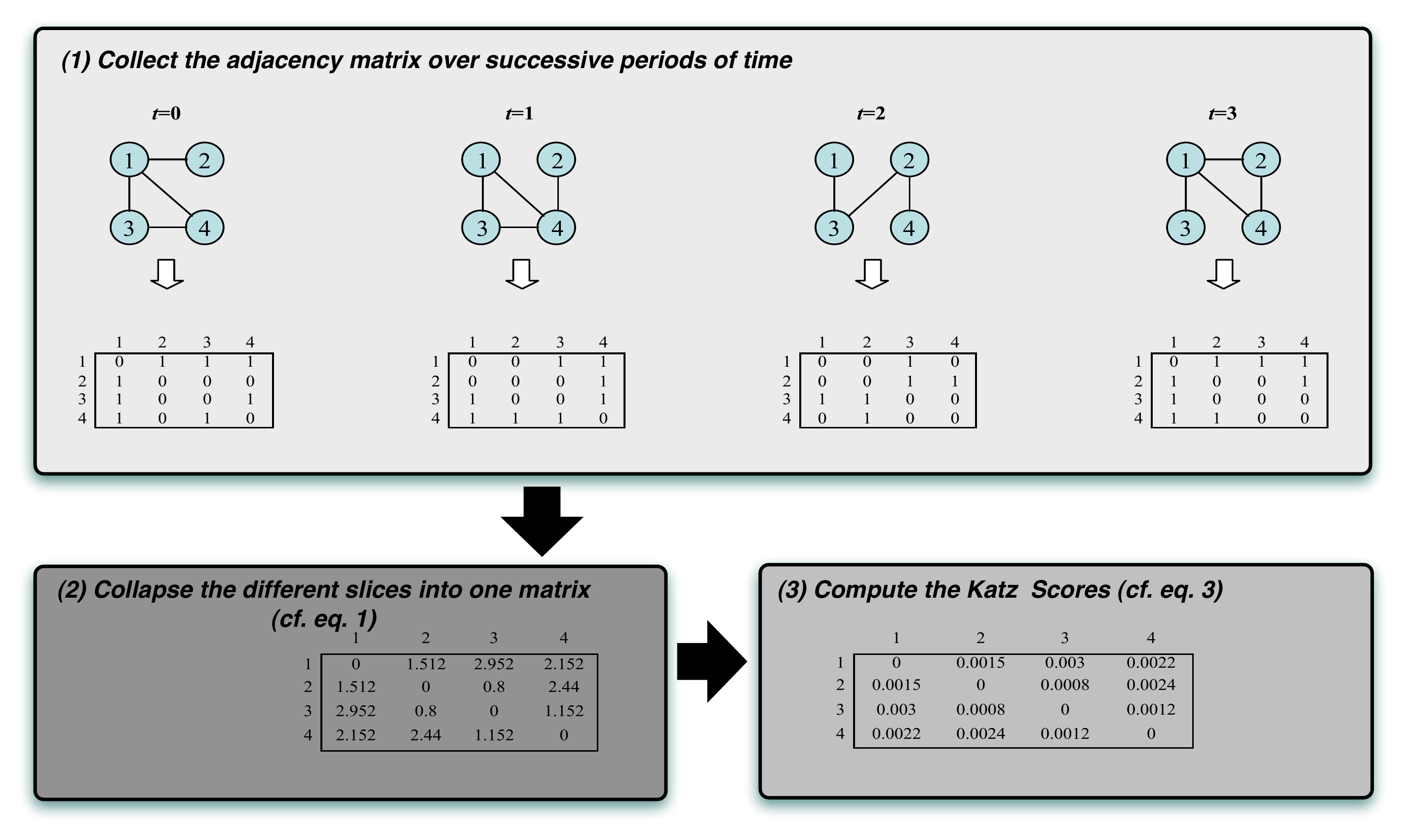}
    \caption{Example of the matrix $\mathbf{S}$ computation}
    \label{Zayani0}
\end{figure}

The computation of the similarity scores is modeled through two
distinct steps. First, we store the inter-contact between nodes in a
tensor $\boldsymbol{\mathcal{Z}}$ and reduce it to a matrix
$\mathbf{X}$ called the collapsed tensor. In a second step, we
compute the matrix of similarity scores $\mathbf{S}$ relying on the
matrix $\mathbf{X}$ (cf. Fig. \ref{Zayani0}).

We consider that the data is collected into the tensor
$\boldsymbol{\mathcal{Z}}$. The slice $\mathbf{Z_{p}}(i, j)$
describes the status of a link between a node $i$ and a node $j$
during a time period $[(p-1) \cdot t,p \cdot t[$ ($p$>0) where
$\mathbf{Z}_{p}(i, j)$ is 1 if the link exists during this period
and 0 otherwise. The tensor is formed by a succession of adjacency
matrices $\mathbf{Z_{1}}$ to $\mathbf{Z_{T}}$ where the subscript
letters designate the observed period. To collapse the data into one
matrix as done in \cite{Acar2009,Dunlavy2011}, we choose to compute
the collapsed weighted tensor (which is the most efficient way to
collapse the data as shown in \cite{Acar2009} and
\cite{Dunlavy2011}). The links structure is considered over time and
the more recent the adjacency matrix is, the more weighted the
structure gets. The collapsed weighted tensor is computed as
following:

\begin{equation}
    \mathbf{X}(i,j)=\sum_{p=1}^{T} (1-\theta)^{T-t}\ \mathbf{Z_{p}}(i,j)
    \label{eq1}
\end{equation}
where the matrix $\mathbf{X}$ is the collapsed weighted tensor of
$\boldsymbol{\mathcal{Z}}$, and $\theta$ is a parameter used to
adjust the weight of recentness and is between 0 and 1.

As Katz measure quantifies the network proximity between two nodes
and given that there are ``social relationships" between nodes in
networks with intermittent connections, it is challenging to exploit
this measure and to apply it on the collected data. Therefore, the
Katz score of a link between a node $i$ and a node $j$ as given by
\cite{Katz1953}:

\begin{equation}
    \mathbf{S}(i,j)=\sum_{\ell=1}^{+\infty} \beta^{\ell} P_{\left \langle \ell  \right \rangle}(i,j)
    \label{eq2}
\end{equation}
where $\beta$ is a user defined parameter strictly superior to zero,
$\beta^{\ell}$ is the weight of a $\ell$ hops path length and
$P_{\left \langle \ell  \right \rangle}(i,j)$ represents the number
of paths of length $\ell$ that join the node $i$ to the node $j$.

It is clear that the longer the path is, the lower the weight gets.
There is also another formulation to compute Katz scores by means of
collapsed weighted tensor as detailed previously. We quantify the
proximity between nodes relying on the paths that separate a pair of
nodes and the weights of the links that form these paths. Then, the
score matrix $\mathbf{S}$ can be rewritten as:

\begin{equation}
    \mathbf{S}=\sum_{\ell=1}^{+\infty} \beta^{\ell} \cdot \mathbf{X}^{\ell}=(\mathbf{I}-\beta \cdot \mathbf{X})^{-1}-\mathbf{I}
    \label{eq3}
\end{equation}
Where $\mathbf{I}$ is the identity matrix and $\mathbf{X}$ is the
collapsed weighted tensor obtained.

We depict as previously mentioned in Fig. \ref{Zayani0} an example
which details the two major steps described before. We take into
consideration a network consisting of 4 nodes and having a dynamic
topology over 4 time periods and we highlight how similarity scores
are obtained. The parameters $\theta$ and $\beta$ are respectively
set to 0.2 and 0.001 for the example and later for the simulations.
We have looked after the values to choose for these two parameters
through several simulations and we have found that such a setting
make possible the convergence of the Katz measure as explained in
\cite{Franceschet2011}. In this example, we assume that all nodes
have the full knowledge of the network structure.

\section{Performance Evaluation and Simulation Results}
To evaluate how efficient is the tensor-based link prediction in
intermittently connected wireless networks, we consider two real
traces. In the following, we firstly present the traces used for the
link prediction evaluation. Then, we expose the corresponding
results, analyze the effectiveness of the prediction method and
compare its performance to those of well-known link prediction
metrics proposed in the literature.

\subsection{Simulation Traces}
We consider two real traces to evaluate the link prediction
approach. We exploit them to construct the tensor by generating
adjacency matrices for several tracking periods. For each case, we
track the required statistics about network topology within $T$
periods. We also consider the adjacency matrix corresponding to the
period $T$+1 as a benchmark to evaluate Katz scores matrix. We
detail, in the following, the used traces.
\begin{itemize}
\item \textbf{First Trace: Dartmouth Campus trace:}
we choose the trace of 01/05/06 \cite{Dartmouth} and construct the
tensor slices relying on SYSLOG traces between 8 a.m. and 3 p.m. (7
hours). The number of nodes is 1018 and the number of locations
(i.e. access points) is 128.
\item \textbf{Second Trace: MIT Campus trace:}
we focus on the trace of 07/23/02 \cite{Balazinska2003} and consider
also the events between 8 a.m. and 3 p.m. to build up the tensor.
The number of nodes is 646 and the number of locations (i.e. access
points) is 174.
\end{itemize}
For each scenario, we generate adjacency matrices corresponding to a
different tracking periods $t$: 5, 10, 30 and 60 minutes. To record
the network statistics over 7 hours, the tensor has respectively a
number of slices $T$ equal to 84, 42, 14 and 7 slices (for the case
where $t$=5 minutes, it is necessary to have 84 periods to cover 7
hours). We take into account both centralized and distributed cases
for the computation of scores.
\begin{itemize}
\item \textbf{The Centralized Computation:}
the centralized way assumes that there is a central entity which has
full knowledge of the network structure at each period and applies
Katz measure to the global adjacency matrices.
\item \textbf{The Distributed Computation:}
each node has a limited knowledge of the network structure. We
assume that a node is aware of its two-hop neighborhood. Hence,
computation of Katz measures is performed on a
local-information-basis.
\end{itemize}

\subsection{Performance Analysis}
As described in the previous section, we apply the link prediction
method to the traces with considering different tensor slice periods
in both centralized and distributed cases. In order to assess the
efficiency of this method, we consider several link prediction
scenarios (according to the trace, the tensor slice period and the
scores computation way) and we use different evaluation techniques.
We detail in the following the results obtained for the evaluation
and analyze the link prediction efficiency. Then, we compare the
performance of the proposed framework to those of major link
prediction metrics in order to justify the use of the Katz measure.

\subsubsection{Evaluation of the link prediction technique}
\begin{figure*}[!tb]
  \centering
  \subfigure[5 minutes tensor slice period]{
    \label{fig1}
    \includegraphics[width=0.15\textwidth,angle=270]{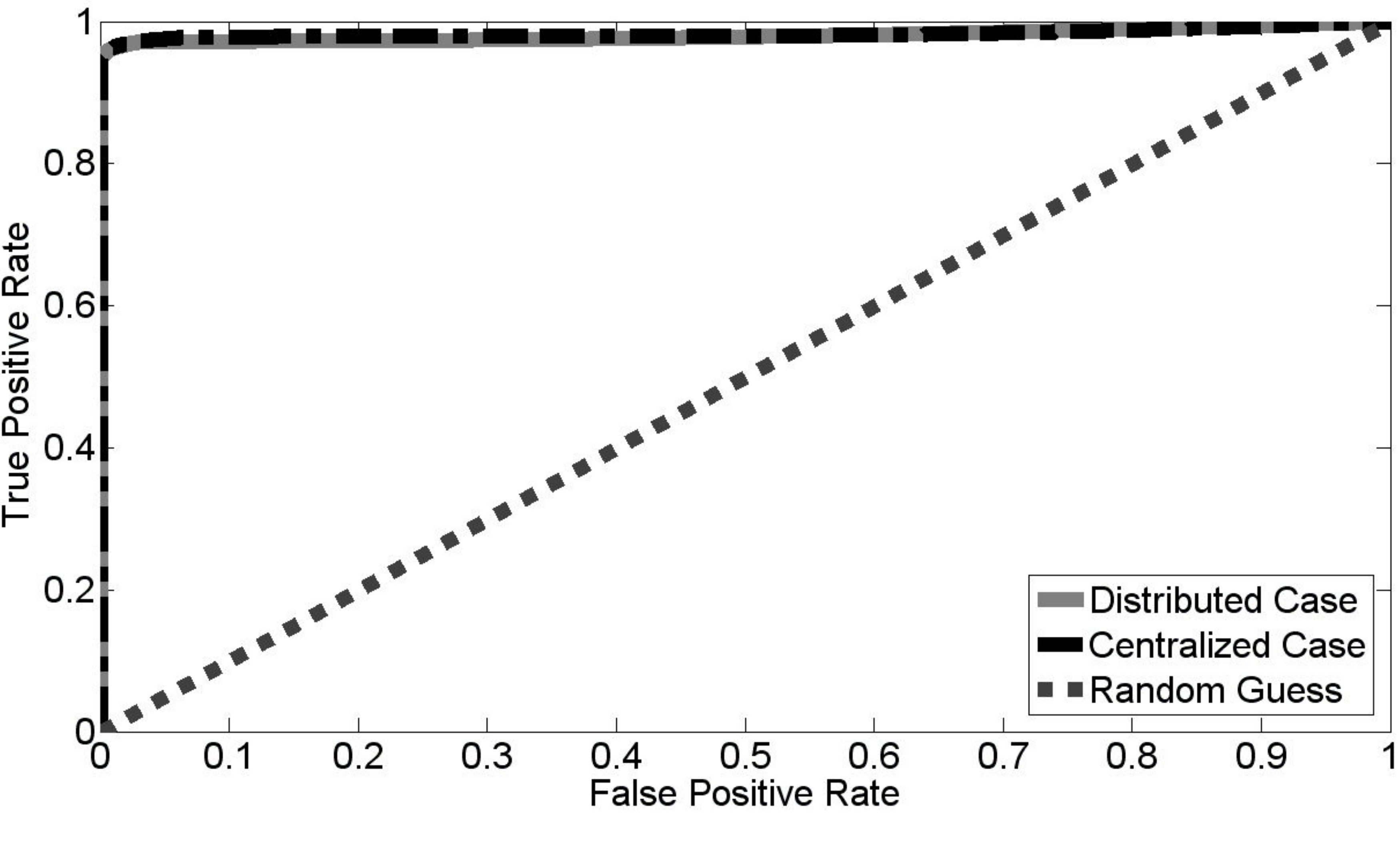}
  }\hspace{1cm}
  \subfigure[10 minutes tensor slice period]{
    \label{fig2}
    \includegraphics[width=0.15\textwidth,angle=270]{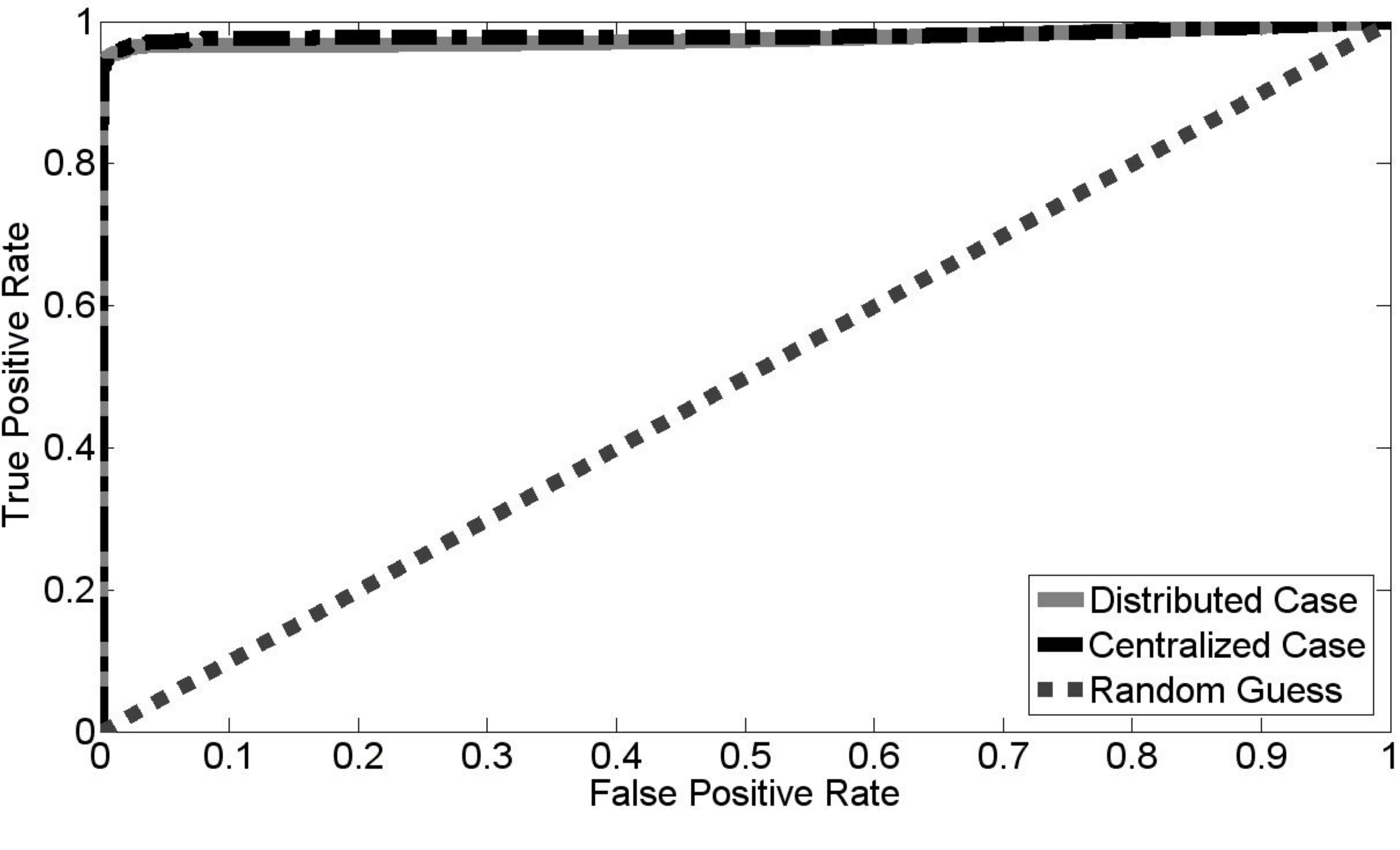}
  }\\
 \subfigure[30 minutes tensor slice period]{
    \label{fig3}
    \includegraphics[width=0.15\textwidth,angle=270]{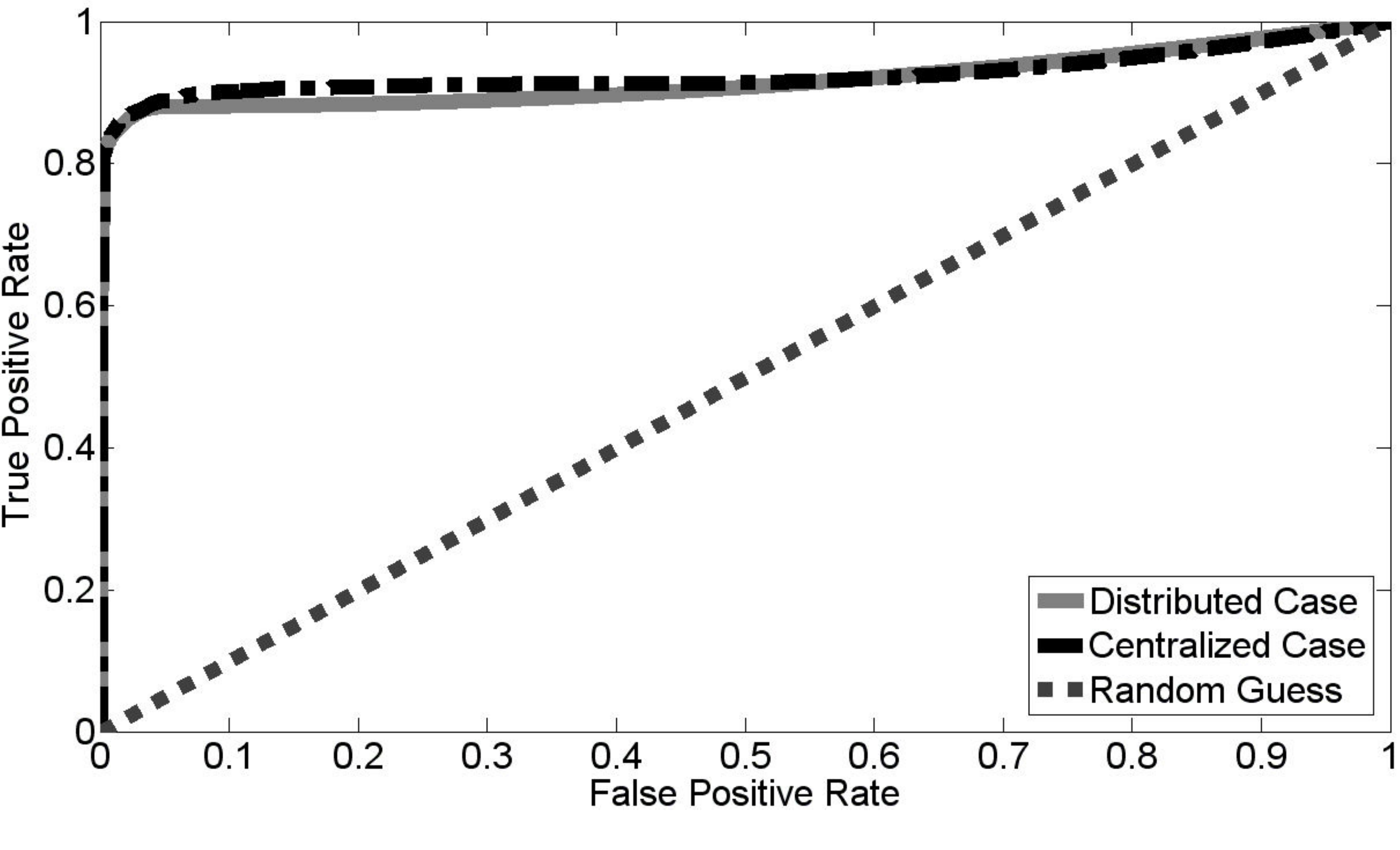}
  }\hspace{1cm}
  \subfigure[60 minutes tensor slice period]{
     \label{fig4}
     \includegraphics[width=0.15\textwidth,angle=270]{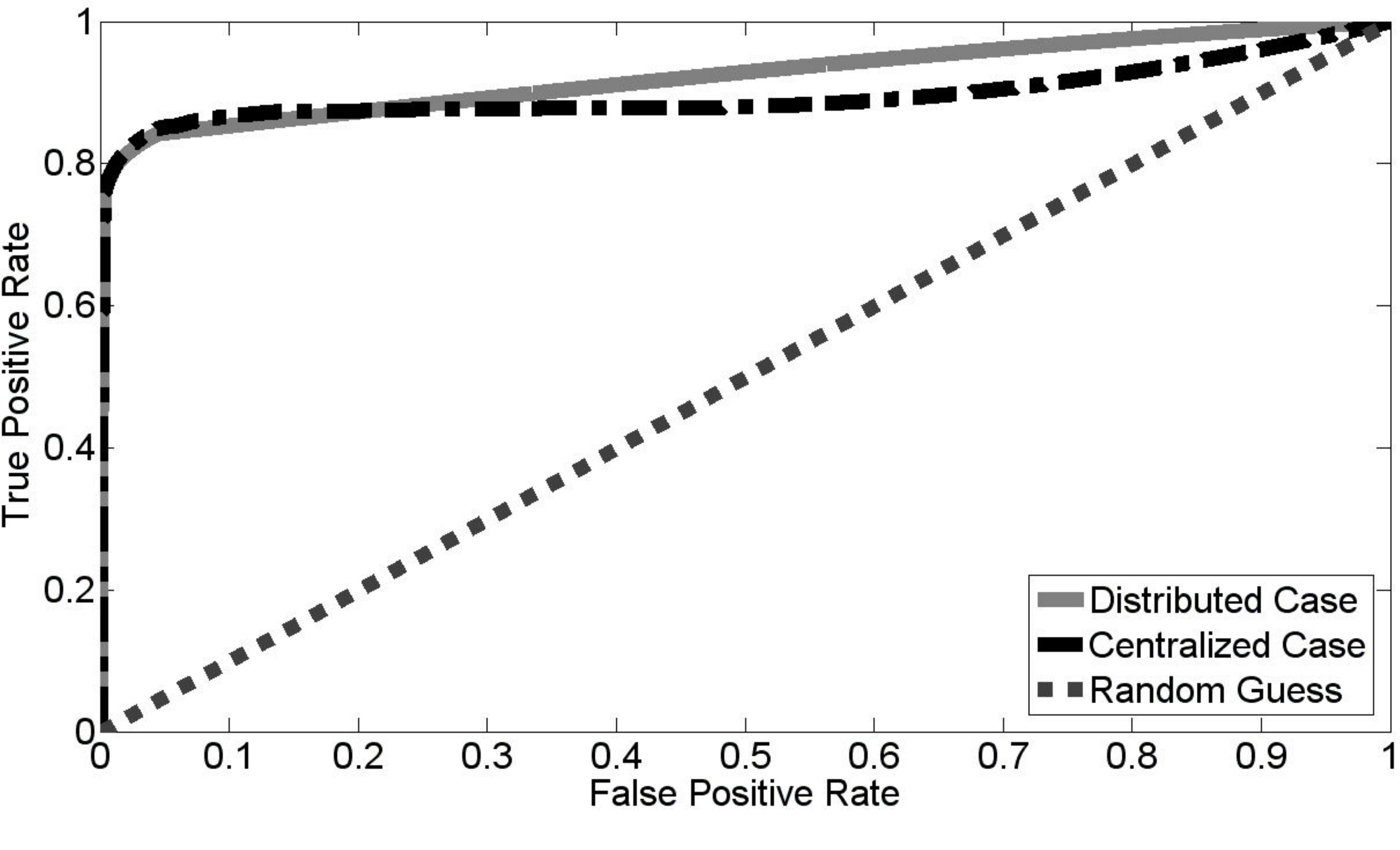}
  }
  \caption{ROC Curves for different prediction cases applied on Dartmouth Campus trace}
  \label{ROC_Dartmouth}
\end{figure*}

To evaluate the efficiency of our proposal, we plot the ROC curves
(Receiver Operating Characteristic curves) \cite{FAWCETT2006}. In
Fig. \ref{ROC_Dartmouth}, we depict the ROC curves obtained after
performing prediction on the Dartmouth Campus trace and for
different tensor slice times. Also, adapted metrics are used in
order to weigh the performance of the proposed link prediction
technique. To this end, we compute the Area Under the ROC Curve
metric (AUC metric) \cite{FAWCETT2006} which could be considered as
a good performance indicator in our case. The AUC metric of each
scenario is determined from the corresponding ROC curve. Moreover,
we consider the top scores ratio metric at $T$+1. To determine this
metric, we compute the accurate number of links identified through
the link prediction technique. We list, for each considered time
period, the number of existing links at period $T$+1, which we call
$L$. Then, we extract the links having the $L$ highest scores and
determine the number of existing links in both sets. The evaluation
metrics are computed for all traces with different tensor slice
periods in both distributed and centralized scenarios. The results
corresponding to all links prediction are listed in Table
\ref{table2} (Dartmouth Campus trace)and Table \ref{table3} (MIT
Campus trace).

\begin{table}[!t]
\renewcommand{\arraystretch}{1}
\caption{Evaluation metrics for the prediction of all links applied
on Dartmouth Campus trace} \label{table2} \centering
\scalebox{0.55}{
\begin{tabular}{|l|c|c|}
\hline
\backslashbox{\bfseries Prediction Cases}{\bfseries Metrics} & \bfseries AUC & \bfseries Top Scores Ratio at $T$+1\\
\hline\hline
Distributed Case and $t$=5 mins & 0.9932 & 93.70\%\\
Centralized Case and $t$=5 mins & 0.9905 & 93.61\%\\
\hline
Distributed Case and $t$=10 mins & 0.9915 & 90.26\% \\
Centralized Case and $t$=10 mins & 0.9883 & 90.19\% \\
\hline
Distributed Case and $t$=30 mins & 0.9813 & 82.31\% \\
Centralized Case and $t$=30 mins & 0.9764 & 82.56\% \\
\hline
Distributed Case and $t$=60 mins & 0.9687 & 76.10\% \\
Centralized Case and $t$=60 mins & 0.9636 & 75.94\% \\
\hline
\end{tabular}}

\end{table}

\begin{table}[!t]
\renewcommand{\arraystretch}{1}
\caption{Evaluation metrics for the prediction of all links applied
on MIT Campus trace} \label{table3} \centering \scalebox{0.55}{
\begin{tabular}{|l|c|c|}
\hline
\backslashbox{\bfseries Prediction Cases}{\bfseries Metrics} & \bfseries AUC & \bfseries Top Scores Ratio at $T$+1\\
\hline\hline
Distributed Case and $t$=5 mins & 0.9907 & 91.48\%\\
Centralized Case and $t$=5 mins & 0.9929 & 91.48\%\\
\hline
Distributed Case and $t$=10 mins & 0.9797 & 85.18\% \\
Centralized Case and $t$=10 mins & 0.9809 & 85.14\% \\
\hline
Distributed Case and $t$=30 mins & 0.9589 & 73.31\% \\
Centralized Case and $t$=30 mins & 0.9578 & 73.76\% \\
\hline
Distributed Case and $t$=60 mins & 0.9328 & 64.54\% \\
Centralized Case and $t$=60 mins & 0.9325 & 64.54\% \\
\hline
\end{tabular}}
\end{table}

\begin{table}[t]
\renewcommand{\arraystretch}{1}
\caption{Table of confusion of a binary prediction technique}
\label{table_confusion} \centering \scalebox{0.55}{
\begin{tabular}{|l|P{0.2}|P{0.2}|}
\hline \backslashbox{Prediction outcome} {Actual
value}& Positive & Negative\\
\hline
Positive & True Positive ($TP$) & False Positive ($FP$)\\
\hline
Negative & False Negative ($FN$) & True Negative ($TN$)\\
\hline
\end{tabular}
}
\end{table}

We first note that, in Fig. \ref{ROC_Dartmouth} and for all
scenarios, the prediction of all links is quite efficient, compared
to the random guess (the curve's bends are at the upper left
corner). We obtain similar ROC curves with the MIT Campus traces (we
do not present them due to space limitations). Moreover, we remark,
based on the high values of AUC metric (over than 0.9) and top
scores ratio obtained at $T$+1, that the prediction method is
efficient in predicting future links (for the period $T$+1). We also
note that prediction is better when the tensor slice periods are
shorter. This observation is obvious for two reasons. On the one
hand, with a low tensor slice time, the probability of tracking a
short and occasional contact between two nodes is not likely. On the
other hand, recording four hours of statistics requires 84 adjacency
matrices of 5-minute periods instead of 7 matrices for 60-minute
periods case. Thus, tracking a short contact between two nodes has
less influence when the tensor slices are more numerous.

Regarding the comparison between the two ways of computing the Katz
scores, we observe that the centralized and distributed matrix of
scores computation achieve similar performances. In fact, the
similarity is higher when the paths considered between a pair of
nodes are short. Thereby, paths that have more than two hops have
weaker scores and so are less weighted compared to shorter ones. The
distributed case assumes that each node knows its neighbors at most
at two hops. That is why distributed scores computation presents
performances which are so similar to the centralized ones.

\subsubsection{Prediction Performance Comparison between the
Tensor-Based Technique and Well-Known Link Prediction Metrics}

We aim through this subsection to compare our proposal to another
similar approaches (we use the distributed design of our framework
to compute the Katz scores). To propose a comprehensive comparison,
we also propose to evaluate the prediction efficiency of well-known
prediction metrics presented in the literature. On the one hand, we
consider behavioral-based link prediction metrics as the similarity
metric of Thakur et al. \cite{Thakur2010} and two metrics expressing
mobile homophily proposed by Wang et al. in \cite{Wang2011}: the
spatial cosine similarity and the co-location rate. On the other
hand, we take two link prediction metrics based on measuring the
degree of proximity as the Katz measure: they are the Adamic-Adar
measure \cite{Adamic2003} and the Jaccard's coefficient
\cite{Jaccard1901,Salton1986}.

To assess the efficiency of each link prediction metric, we consider
these evaluation measures:
\begin{itemize}
\item \textbf{Top Scores Ratio in the period $T$+1 (TSR):} to determine this metric, we compute
the percentage of occurring links identified through the link
prediction technique. We list the number of existing links (at
period $T$+1 or during the periods coming after the period $T$)
which we call $L$. Then, we extract the pair of nodes having the $L$
highest scores and determine the percentage of links involved in
both sets. existing links in both sets.
\item \textbf{Accuracy (ACC):} this measure is defined in \cite{FAWCETT2006} as
the ratio of correct prediction (true positive and true negative
predictions) over all predictions (true positive, true negative,
false positive and false negative predictions). In other words, it
is computed by the ratio $\frac{TP+TN}{TP+FP+TN+FN}$ (see Table
\ref{table_confusion}). We identify for each scenario the maximum
value of the accuracy which indicates the degree of precision that
can reach each prediction metric.
\item \textbf{Precision or Positive Predictive Value (PPV):}
it represents to the proportion of links with positive prediction
(occurring in the future) which are correctly identified
\cite{FAWCETT2006}. Based on Table \ref{table_confusion}, the
precision is equal to $\frac{TP}{TP+FP}$. This value is determined
according to the deduced accuracy value.
\item \textbf{Recall or True Positive Rate (TPR):} it quantifies
the ratio of correctly identified links over the occurring links in
the future \cite{FAWCETT2006}. Referring to Table
\ref{table_confusion}, the recall is defined by the expression
$\frac{TP}{TP+FN}$. This value is also computed according to the
retained accuracy value.
\item \textbf{F-measure or balanced F1 score:} the F-measure \cite{vanRijsbergen1979} is the harmonic mean of
precision and recall. The F-measure is expressed by
$2.\frac{precision.recall}{precision+recall}$. The higher the
F-measure is, the better the tradeoff of precision and recall gets
and the more efficient the prediction metric is.
\end{itemize}

The evaluation metrics are computed for all traces with different
tracking periods lengths (5, 10, 30 and 60 minutes). For each trace,
we track the network topology from 8 a.m. to 4 p.m. We divide, as
previously, the historical into $T$ periods and we focus on
predicting the links occurring in the period $T$+1. Regarding the
Dartmouth Campus trace, the results are reported in Table
\ref{table_Dartmouth}. For the MIT Campus trace, the prediction
results are listed in Table \ref{table_MIT}.

\begin{table}[!tb]
\renewcommand{\arraystretch}{1}
\caption{Evaluation metrics for the prediction applied on the
Dartmouth Campus trace} \label{table_Dartmouth} \centering
\scalebox{0.55}{
\begin{tabular}{|P{0.07}|P{0.17}|P{0.078}|P{0.078}|P{0.078}|P{0.078}|P{0.078}|}
\hline
\bfseries Period length & \bfseries Prediction Score & \bfseries TSR in $T$+1 & \bfseries Accuracy & \bfseries Precision (PPV) & \bfseries Recall (TPR) & \bfseries F-measure\\

\hline\hline \multirow{6}{2cm}{5 mins ($T$=96)}
 & Thakur's Metric                & 41.39\% & 99.11\% & 36.40\% & 11.57\% & 0.1756 \\  \cline{2-7}
 & Spatial Cosine Sim.            & 66.01\% & 99.45\% & 67.44\% & 63.75\% & 0.6554 \\  \cline{2-7}
 & Co-Location Rate               & 68.96\% & 99.50\% & 73.98\% & 60.71\% & 0.6669 \\  \cline{2-7}
 \hhline{|~|------}
 \hhline{|~|------}
 \hhline{|~|------}
 \hhline{|~|------}
 & Adamic-Adar Meas.              & 83.81\% & 99.74\% & 82.58\% & 85.57\% & 0.8405 \\  \cline{2-7}
 & Jaccard's Coeff.               & 82.54\% & 99.72\% & 81.27\% & 85.08\% & 0.8313 \\  \cline{2-7}
 & Katz Measure                   & 90.88\% & 99.86\% & 90.59\% & 91.87\% & 0.9123 \\  \cline{2-7}
\hline\hline \multirow{6}{2cm}{10 mins ($T$=48)}
 & Thakur's Metric                & 43.29\% & 99.10\% & 37.31\% & 11.15\% & 0.1717 \\  \cline{2-7}
 & Spatial Cosine Sim.            & 66.71\% & 99.45\% & 68.52\% & 62.99\% & 0.6564 \\  \cline{2-7}
 & Co-Location Rate               & 68.78\% & 99.49\% & 71.50\% & 65.63\% & 0.6844 \\  \cline{2-7}
 \hhline{|~|------}
 \hhline{|~|------}
 \hhline{|~|------}
 \hhline{|~|------}
 & Adamic-Adar Meas.              & 81.01\% & 99.68\% & 78.87\% & 84.00\% & 0.8135 \\  \cline{2-7}
 & Jaccard's Coeff.               & 79.75\% & 99.66\% & 78.04\% & 82.83\% & 0.8036 \\  \cline{2-7}
 & Katz Measure                   & 86.39\% & 99.78\% & 89.75\% & 82.94\% & 0.8621 \\  \cline{2-7}
\hline\hline \multirow{6}{2cm}{30 mins ($T$=16)}
 & Thakur's Metric                & 45.18\% & 99.06\% & 39.08\% & 10.83\% & 0.1696 \\  \cline{2-7}
 & Spatial Cosine Sim.            & 67.35\% & 99.42\% & 67.60\% & 67.00\% & 0.6730 \\  \cline{2-7}
 & Co-Location Rate               & 67.78\% & 99.45\% & 72.47\% & 61.33\% & 0.6644 \\  \cline{2-7}
 \hhline{|~|------}
 \hhline{|~|------}
 \hhline{|~|------}
 \hhline{|~|------}
 & Adamic-Adar Meas.              & 71.82\% & 99.50\% & 71.25\% & 73.86\% & 0.7253 \\  \cline{2-7}
 & Jaccard's Coeff.               & 71.34\% & 99.50\% & 72.63\% & 69.65\% & 0.7111 \\  \cline{2-7}
 & Katz Measure                   & 79.83\% & 99.64\% & 80.09\% & 79.48\% & 0.7978 \\  \cline{2-7}
\hline\hline \multirow{6}{2cm}{60 mins ($T$=8)}
 & Thakur's Metric                & 46.39\% & 99.04\% & 41.39\% & 10.61\% & 0.1689 \\  \cline{2-7}
 & Spatial Cosine Sim.            & 67.55\% & 99.40\% & 68.51\% & 65.70\% & 0.6708 \\  \cline{2-7}
 & Co-Location Rate               & 68.11\% & 99.42\% & 72.21\% & 60.31\% & 0.6573 \\  \cline{2-7}
 \hhline{|~|------}
 \hhline{|~|------}
 \hhline{|~|------}
 \hhline{|~|------}
 & Adamic-Adar Meas.              & 65.98\% & 99.38\% & 69.73\% & 57.42\% & 0.6298 \\  \cline{2-7}
 & Jaccard's Coeff.               & 67.00\% & 99.47\% & 68.31\% & 64.53\% & 0.6637 \\  \cline{2-7}
 & Katz Measure                   & 74.09\% & 99.53\% & 75.33\% & 72.84\% & 0.7406 \\  \cline{2-7}
\hline
\end{tabular}}
\end{table}

\begin{table}[!tb]
\renewcommand{\arraystretch}{1}
\caption{Evaluation metrics for the prediction applied on the MIT
Campus trace} \label{table_MIT} \centering \scalebox{0.55}{
\begin{tabular}{|P{0.07}|P{0.17}|P{0.078}|P{0.078}|P{0.078}|P{0.078}|P{0.078}|}
\hline
\bfseries Period length & \bfseries Prediction Score & \bfseries TSR in $T$+1 & \bfseries Accuracy & \bfseries Precision (PPV) & \bfseries Recall (TPR) & \bfseries F-measure\\

\hline\hline \multirow{6}{2cm}{5 mins ($T$=96)}
 & Thakur's Metric                & 58.22\% & 99.29\% & 65.58\% & 44.96\% & 0.5335 \\  \cline{2-7}
 & Spatial Cosine Sim.            & 60.87\% & 99.34\% & 72.56\% & 44.17\% & 0.5491 \\  \cline{2-7}
 & Co-Location Rate               & 69.35\% & 99.49\% & 77.79\% & 60.71\% & 0.6820 \\  \cline{2-7}
 \hhline{|~|------}
 \hhline{|~|------}
 \hhline{|~|------}
 \hhline{|~|------}
 & Adamic-Adar Meas.              & 84.20\% & 99.72\% & 84.22\% & 84.36\% & 0.8429 \\  \cline{2-7}
 & Jaccard's Coeff.               & 82.18\% & 99.68\% & 83.11\% & 81.12\% & 0.8210 \\  \cline{2-7}
 & Katz Measure                   & 90.14\% & 99.86\% & 95.29\% & 89.02\% & 0.9205 \\  \cline{2-7}
\hline\hline \multirow{6}{2cm}{10 mins ($T$=48)}
 & Thakur's Metric                & 57.70\% & 99.27\% & 65.25\% & 44.58\% & 0.5569 \\  \cline{2-7}
 & Spatial Cosine Sim.            & 60.50\% & 99.32\% & 72.56\% & 43.18\% & 0.5414 \\  \cline{2-7}
 & Co-Location Rate               & 68.74\% & 99.46\% & 76.50\% & 60.08\% & 0.6730 \\  \cline{2-7}
 \hhline{|~|------}
 \hhline{|~|------}
 \hhline{|~|------}
 \hhline{|~|------}
 & Adamic-Adar Meas.              & 80.04\% & 99.63\% & 79.31\% & 80.87\% & 0.8008 \\  \cline{2-7}
 & Jaccard's Coeff.               & 77.97\% & 99.59\% & 80.53\% & 73.77\% & 0.7700 \\  \cline{2-7}
 & Katz Measure                   & 86.83\% & 99.78\% & 86.62\% & 87.25\% & 0.8693 \\  \cline{2-7}
\hline\hline \multirow{6}{2cm}{30 mins ($T$=16)}
 & Thakur's Metric                & 56.73\% & 99.20\% & 62.87\% & 46.14\% & 0.5322 \\  \cline{2-7}
 & Spatial Cosine Sim.            & 59.35\% & 99.26\% & 72.65\% & 40.51\% & 0.5202 \\  \cline{2-7}
 & Co-Location Rate               & 65.03\% & 99.39\% & 80.75\% & 49.93\% & 0.6171 \\  \cline{2-7}
 \hhline{|~|------}
 \hhline{|~|------}
 \hhline{|~|------}
 \hhline{|~|------}
 & Adamic-Adar Meas.              & 67.07\% & 99.35\% & 67.77\% & 64.55\% & 0.6572 \\  \cline{2-7}
 & Jaccard's Coeff.               & 66.34\% & 99.39\% & 78.56\% & 51.97\% & 0.6214 \\  \cline{2-7}
 & Katz Measure                   & 72.85\% & 99.47\% & 88.30\% & 53.86\% & 0.7279 \\  \cline{2-7}
\hline\hline \multirow{6}{2cm}{60 mins ($T$=8)}
 & Thakur's Metric                & 55.70\% & 99.08\% & 63.51\% & 41.85\% & 0.5045 \\  \cline{2-7}
 & Spatial Cosine Sim.            & 57.57\% & 99.14\% & 72.82\% & 37.22\% & 0.4926 \\  \cline{2-7}
 & Co-Location Rate               & 61.71\% & 99.24\% & 77.89\% & 45.10\% & 0.5712 \\  \cline{2-7}
 \hhline{|~|------}
 \hhline{|~|------}
 \hhline{|~|------}
 \hhline{|~|------}
 & Adamic-Adar Meas.              & 59.13\% & 99.14\% & 57.52\% & 65.48\% & 0.6124 \\  \cline{2-7}
 & Jaccard's Coeff.               & 58.95\% & 99.22\% & 71.27\% & 45.73\% & 0.5571 \\  \cline{2-7}
 & Katz Measure                   & 61.00\% & 99.28\% & 74.90\% & 50.09\% & 0.6003 \\  \cline{2-7}
\hline
\end{tabular}}
\end{table}

The results obtained enable us to attest that the use of the Katz
measure has been one of the best choices to perform prediction
through the tensor-based technique. Using this metric achieves
better performance than those of the other link prediction metrics
proposed in the literature. Hence, the Katz measure is the best
metric that we can use to perform link prediction.

The prediction made through the Katz measure achieves better
performance than those of mobility homophily metrics and Thakur et
al.'s similarity. Indeed, our framework quantifies the similarity of
nodes based on their encounters and geographical closeness. In other
words, the proposed prediction method cares about contacts (or
closenesses) at (around) the same location and at the same time.
Meanwhile, the mobility homophily metrics and Thakur et al.'s
similarity are defined as an association metric. Hence, they measure
the degree of similarity of behaviors of two mobile nodes without
necessarily seeking if they are in the same location at the same
time. Regarding the comparison with the other network proximity
metrics, the Katz measure quantifies better the behavior similarity
of two nodes as it takes into consideration only the paths that
separate them. Meanwhile, the Adamic-Adar metric and the Jaccard's
coefficient are dependent respectively on the degree of common
neighbors between two nodes and the size of the intersection of the
neighbors of two nodes. These latter metrics express similarity
based on common neighbors of two nodes but don't seek if a link is
occurring between them. This criterion highly influences the value
of Katz measure and make it more precise.

\section{Conclusion}
Human mobility patterns are mostly driven by social intentions and
correlations appear in the behavior of people forming the network.
These similarities highly govern the mobility of people and then
directly influence the structure of the network. The knowledge about
the behavior of nodes greatly helps in improving the design of
communication protocols. Intuitively, two nodes that follow the same
social intentions over time promote the occurrence of a link in the
immediate future.

In this paper, we presented a link prediction technique inspired by
data mining and exploit it in the context of human-centered wireless
networks. Through the link prediction evaluation, we have obtained
relevant results that attest the efficiency of our contribution and
agree
with some findings referred in the literature. 
%
%

Good link prediction offers the possibility to further improve
opportunistic packet forwarding strategies by making better
decisions in order to enhance the delivery rate or limiting latency.
Therefore, it will be relevant to supply some routing protocols with
prediction information and to assess the contribution of our
approach in enhancing the performance of the network especially as
we propose an efficient distributed version of the prediction
method. The proposed technique also motivates us to inquire into
future enhancements as a more precise tracking of the behavior of
nodes and a more efficient similarity computation.

\bibliographystyle{ieeetran}
\bibliography{Biblio}

\end{document}